\documentclass[aps,prl,amsmath,amssymb,twocolumn]{revtex4}
\usepackage[latin1]{inputenc}
\usepackage{graphicx}

\begin{document}

\title{Population and Phase Coherence during the Growth of an Elongated Bose-Einstein Condensate}
\author{M.~Hugbart}
\author{J.~A.~Retter}
\author{A.~F.~Var\'on}
\author{P.~Bouyer}\email{philippe.bouyer@iota.u-psud.fr}
\author{A.~Aspect}

\affiliation{Laboratoire Charles Fabry de l'Institut d'Optique,
Centre National de la Recherche Scientifique et Université Paris
Sud 11, Batiment 503, Centre Scientifique F91403 ORSAY CEDEX,
France.}
\author{M.~J.~Davis}
\affiliation{ARC Centre of Excellence for Quantum-Atom Optics,
School of Physical Sciences,  University of Queensland, Brisbane,
QLD 4072, Australia.}
\date{\today}
\begin{abstract}
We study the growth of an elongated phase-fluctuating condensate
from a non-equilibrium thermal cloud obtained by shock-cooling. We
compare the growth of the condensate with numerical simulations,
revealing a time delay and a reduction in the growth rate which we
attribute to phase fluctuations.  We measure the phase coherence
using momentum Bragg spectroscopy, and thereby observe the
evolution of the phase coherence as a function of time.  Combining
the phase coherence results with the numerical simulations, we
suggest a simple model for the reduction of the growth rate based
on the reduction of bosonic stimulation due to phase fluctuations
and obtain improved agreement between theory and experiment.
\end{abstract}
\maketitle

%\section{Introduction}
The non-equilibrium path to Bose-Einstein condensation is a
complex process in which atoms accumulate in the ground state of
the system and long-range phase coherence develops, resulting in a
strong suppression of density fluctuations and a uniform phase.
The kinetics of condensate formation has long been a subject of
theoretical study, giving rise to a number of conflicting
predictions (see \cite{Sto99} for a review). Quantitative theories
have been formulated to model the condensate formation process in
a harmonic trapping potential \cite{Gar97, Bil00}.  However, a
limitation of these models is that the condensate is assumed to
grow adiabatically in its phase coherent ground state. On the
other hand, for a homogeneous system, Kagan \textit{et
al.}~\cite{Kag92} proposed the appearance of a quasi-condensate
with strong phase fluctuations which die out on a time scale that
increases with the size of the system.   This homogeneous system
description is also relevant to condensate growth in hydrodynamic
clouds, where the trapping potential can be neglected
\cite{Svi01}. Condensates in highly-elongated traps, which can
often be treated using the local density approximation, are
expected to have properties close to the homogeneous case
\cite{Pet01}. In addition, the axially hydrodynamic regime is
easily attainable in such traps \cite{Ger04L}.

Experimentally, the problem of condensate formation has been
approached by shock-cooling \cite{Mie98,Koh02,Shv02} in harmonic
traps: starting from a thermal cloud just above the transition
temperature,  rapid removal of the most energetic atoms from the
trap results in an over-saturated thermal cloud. Subsequent
thermalization leads to the growth of the condensate. Measurements
of the growth of the condensed fraction in traps significantly
less elongated than ours \cite{Mie98,Koh02} have obtained good
quantitative agreement with theory \cite{Koh02,Dav02}, but these
experiments did not give access to the phase coherence of the
growing condensate. Although the two-step growth curve reported in
Ref.~\cite{Koh02}, and the growth of non-equilibrium,
phase-fluctuating condensates from hydrodynamic clouds in
Ref.~\cite{Shv02} support the existence of a quasi-condensate
during the initial stage of condensate formation as proposed in
Ref.~\cite{Kag92}, there exists to our knowledge no quantitative
experimental study of the development of phase coherence.

\begin{figure}[b!]
\includegraphics[width=3in]{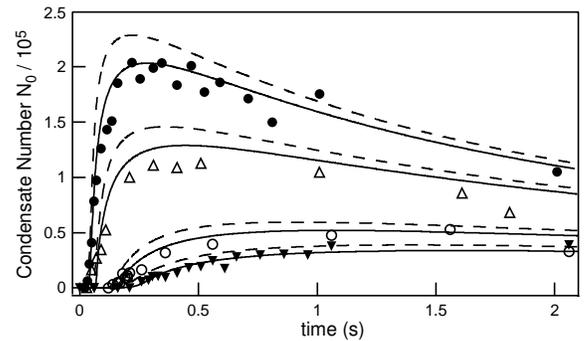}
\caption{\label{GrowthCurves}Condensate growth curves for initial
atom numbers: $N_{\rm{i}}/10^5=8.6(9)\bullet$,
$7.2(7)$$\vartriangle$, %$4.4(3)${\tiny{${\blacksquare}$}},
$3.4(4)\circ$ and $2.8(5) \blacktriangledown$ \cite{rescale}. Each
point corresponds to an average over three experimental
realizations. The solid (dashed) lines are theoretical results of
the model of \cite{Dav02}, with (without) the correction for
thermal phase fluctuations, as described in the text.  They have
additional delay times of 10, 20, 50, and 50\,ms respectively. The
decrease in $N_0$ at longer times is due to three-body losses.}
\end{figure}

In this Letter we present an experimental study of the evolution
of both the condensate number and the phase coherence during the
growth of a condensate in a highly elongated trap in the axially
hydrodynamic regime.  We observe that the growth is slower than
that predicted by theory and attribute this to the presence of
phase fluctuations which are expected to be strong in our
experimental regime, even at equilibrium \cite{Pet01}. We then use
Bragg spectroscopy \cite{Ste99,Ric03} to measure the coherence
length during the growth of the condensate. Our observations are
compatible with the scenario of Refs.~\cite{Svi01, Kag92}, where a
non-equilibrium quasi-condensate is created at the onset of
condensation and relaxes rapidly to equilibrium with shape
oscillations.  Finally, we propose a simple modification to the
numerical simulations to phenomenologically account for the effect
of phase fluctuations on the condensate growth rate, and find that
this improves the agreement between theory and experiment.

In our experiment \cite{Des99}, we prepare a thermal cloud of
$^{87}$Rb atoms in the $5S_{1/2}|F=1,m_F=-1\rangle$ state in an
Ioffe-Pritchard trap with final trap frequencies of
$\omega_\bot=2\pi \times 655(4)\,$Hz radially and $\omega_z=2\pi
\times 6.53(1)\,$Hz axially.  Forced radio-frequency (rf)
evaporation proceeds to a frequency $\Delta
\nu_{\rm{rf}}=140$\,kHz above that corresponding to the bottom of
the combined magnetic and gravitational potential, giving an
effective trap depth of $\varepsilon_{\rm{i}} =6~\mu$K, and the rf
knife is held at this value for a time varying from 1\,s to 12\,s.
This ensures thermal equilibrium, and allows us to control the
atom number $N_{\rm{i}}$ in the range 2.7--8.6$\times 10^5$. The
resulting thermal cloud has a temperature $T_{\rm{i}}$ of about
$\varepsilon_{\rm{i}}/10$, just above the transition temperature
$T_{\rm{c}}$, which varies from 400\,nK to 600\,nK depending on
the atom number.

We next shock-cool the cloud by rapidly ramping the rf knife to
$\Delta \nu_{\rm{rf}}=40$\,kHz in 25\,ms, giving a final trap
depth $\varepsilon_{\rm{f}}= 1.5~\mu$K. The relative truncation
rate $\dot{\varepsilon}/\varepsilon_{\rm{f}} = 120$~s$^{-1}$ is
fast compared to the axial trap frequency, but slow compared to
the radial trap frequency. In our elongated geometry this
shock-cooling results in a cloud transversally at equilibrium but
axially out of equilibrium. The cloud tends towards local thermal
equilibrium with a temperature $T<T_{\rm{c}}$, in a time $\sim
3\,\tau_{\rm{coll}}\lesssim 10$ ms \cite{therm} where
$\tau_{\rm{coll}}$ is the collision rate at the centre of the trap
\cite{taucoll}. Since the atom cloud is in the hydrodynamic regime
axially ($\omega_z \tau_{\rm{coll}}\ll 1$) \cite{hydro}, global
equilibrium is reached on a time scale longer than the axial
oscillation period.

In order to study the condensate growth, the cloud is held in the
trap for a further time $t$ after the end of the shock-cooling
ramp, with the trap depth held constant at $\varepsilon_{\rm{f}}$.
We then switch off the trap and image the cloud after a 20\,ms
time-of-flight in order to obtain the total atom number $N$,
temperature $T$ and condensate number $N_0$ \cite{Ger04A, Ger04L}.
By repeating the measurements at different times $t$ for the same
initial conditions, we obtain a growth curve for the condensate
number, as shown in Fig.~\ref{GrowthCurves} for various initial
atom numbers $N_{\rm{i}}$ \cite{rescale}. At $t\simeq 20$\,ms
(depending on initial conditions), the atom number has dropped by
40\% and the temperature is already below $T_{\rm{c}}$, yet the
condensate does not appear until later, with a delay time of
20--200\,ms after the fast ramp.

We have simulated the evaporative cooling and condensate growth
for our experiment based on the model described in \cite{Dav02}
with the additional inclusion of three-body loss for the thermal
cloud \cite{Dav06}. The results are shown in
Fig.~\ref{GrowthCurves} (dashed lines). This method was in good
quantitative agreement with experimental results for a
less-elongated system \cite{Koh02}. There are no free parameters
in our calculations which are based on our measured data
\cite{rescale}, however the results plotted in Fig.~1 (dashed
lines) have had a delay time of 10--50\,ms added to better fit the
experimental results. The observed growth rate is somewhat slower
than the theoretical prediction. However, the simulation is not
necessarily valid for such an elongated system, in the
hydrodynamic regime, nor does it take account of phase
fluctuations. Indeed, our experiment is performed in an elongated
trapping geometry, where temperature-dependent phase-fluctuations
can be present even at equilibrium
\cite{Pet01,Det01,Ric03,Hel03,Hug05}. For our parameters, the
phase-coherence length $L_\phi=15\hbar^2 N_0/16 m k_{\rm{B}}LT$
\cite{Pet01} at equilibrium is smaller than the condensate
half-length $L$ by a factor in the range 4--10, varying inversely
with the condensate atom number.  We will return to this point in
more detail after discussing our experimental study of the phase
coherence.

\begin{figure}[b!]
\includegraphics[width=2.8in]{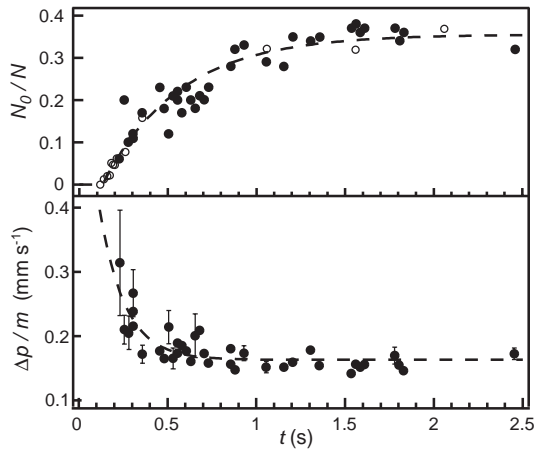}
\caption{\label{fig:Delta_nu}  Condensed fraction $N_0/N$ and
momentum width $\Delta p$ as a function of $t$ for an initial atom
number $N_{\rm{i}}=3.8(3)\times 10^5$.  (Open circles correspond
to the data from Fig.~\ref{GrowthCurves} for
$N_{\rm{i}}=3.4(3)\times 10^5$.) The decreasing momentum width
indicates the growth of the coherence length with time. The dashed
lines are guides to the eye. Some typical error bars are shown.}
\end{figure}
We measure the coherence length of the condensate during its
formation via its momentum distribution, using 4-photon Bragg
Spectroscopy as described in Ref.~\cite{Ric03}. At time $t$ after
the end of the shock-cooling ramp, the magnetic trap is switched
off and after 2\,ms of free expansion a 2\,ms Bragg pulse is
applied. The atoms are imaged after a further 16\,ms
time-of-flight, which allows separation of the diffracted atoms.
The diffracted fraction is measured as a function of $\nu$, the
detuning between the Bragg beams which determines the
velocity-class diffracted, to obtain the momentum spectrum of the
condensate.  We fit a Lorentzian function to the measured spectra
and extract the half-width at half-maximum (HWHM) $\Delta \nu =
2k_{\rm{L}}\Delta p/2\pi m$, where $m$ is the atomic mass,
$k_{\rm{L}}$ the laser wave-vector and $\Delta p$ the HWHM of the
momentum distribution. For each spectrum, further images
(typically 5) are taken without the Bragg pulses, from which the
temperature $T$, condensate number $N_0$, and condensate
half-length $L$ are obtained.

The evolution of the momentum width $\Delta p$ for an initial atom
number $N_{\rm{i}}=3.8(3)\times 10^5$ is presented in
Fig.~\ref{fig:Delta_nu} (lower panel). The corresponding condensed
fractions are shown in the upper panel of Fig.~\ref{fig:Delta_nu}
(filled circles). The momentum width $\Delta p$ clearly decreases
as the condensate fraction grows to equilibrium, indicating that,
as expected, the coherence length grows with time.  Indeed, for a
condensate at equilibrium, we know \cite{Ric03} that the reduction
in coherence length due to thermal phase fluctuations leads to a
broadening of the momentum width, given by \cite{Ger03,Pet01}:
\begin{equation}\label{equ:Dnu}
\Delta p_{\rm{equ}}=\hbar\, \sqrt{\left(
\frac{2.04}{L}\right)^2+\left( \frac {0.65} {L_{\phi}}
\right)^2}.\end{equation} The first term accounts for the
Heisenberg-limited momentum width due to the finite size $L$ of
the condensate and the second term accounts for the presence of
thermal phase fluctuations. The numerical factors account for
integration over the 3D density profile.  We also correct for the
finite ``instrumental width'' of the Bragg spectrometer, by
introducing as in \cite{Ric03} a Gaussian apparatus function of
half-width $w_{\rm{G}}=200$ Hz. This results in a theoretical
momentum width: $\Delta p_{\rm{th}}=\nolinebreak{\Delta
p_{\rm{equ}}}/{2}+\nolinebreak\sqrt{\left({2 \pi
m}/{2k_{\rm{L}}}\right)^2 w_{\rm{G}}^2+\left({\Delta
p_{\rm{equ}}}/{2}\right)^2}$. Equation \ref{equ:Dnu} shows that,
even if the condensate were at equilibrium at each instant during
the growth, we would expect the momentum width to decrease with
time, since both $L_\phi$ and $L$ increase with the condensate
atom number. We can therefore test whether the condensate
coherence follows the density in this way by comparing each
measured momentum width $\Delta p$ with the value $\Delta
p_{\rm{th}}$ calculated for a condensate at equilibrium, using the
parameters $N_0$, $L$ and $T$ measured for each Bragg spectrum.
\begin{figure}[t]
\includegraphics[width=3.1in]{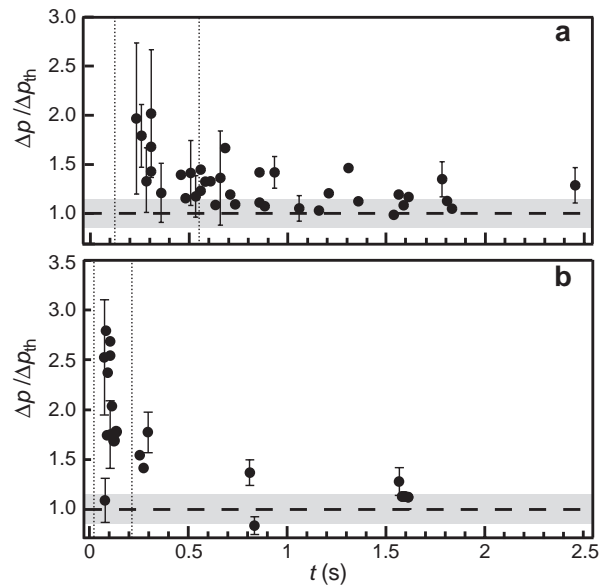}
\caption{\label{fig:Delta_nu_comp} Ratio of measured momentum
width $\Delta p$ to theoretical momentum width $\Delta
p_{\rm{th}}$, calculated for a condensate at equilibrium (see
text) for: (a) $N_{\rm{i}}=3.8(3)\times 10^5$ and (b)
$N_{\rm{i}}=7.2(3)\times 10^5$. (Data shown in (a) corresponds to
Fig.~\ref{fig:Delta_nu}.) The condensate momentum width tends to
the equilibrium value (dashed line) at long times.  Some typical
statistical error bars are shown; the gray band indicates
systematic uncertainties on the equilibrium value. The first
vertical line marks the onset of condensation;  the second
indicates the time at which the condensed fraction reaches
$(1-1/e)$ of its final value.}
\end{figure}
We plot the ratio $\Delta p/\Delta p_{\rm{th}}$ in
Fig.~\ref{fig:Delta_nu_comp} for two different initial atom
numbers. The dashed line at $\Delta p/\Delta p_{\rm{th}}=1$
indicates the value expected for a condensate always at
equilibrium.  Systematic uncertainties of $15\%$ on this
equilibrium value, mainly due to the atom number calibration
($20\%$) and determination of $w_{\rm{G}}$ ($10\%$), are
represented by the gray band. Unambiguously, we observe that the
ratio $\Delta p/\Delta p_{\rm{th}}$ always lies above one (dashed
line), and decreases in time.  This indicates an excess momentum
spread with respect to a condensate at equilibrium during the
growth.  As the condensate reaches equilibrium, the measurement
dispersion decays and the momentum width tends to the predicted
equilibrium value.

To interpret our results, we consider the scenario proposed by
Kagan \textit{et al.}~\cite{Kag92} for condensate growth in a
homogeneous system. The early stages of growth lead to a
quasi-condensate with non-equilibrium, long-range phase
fluctuations in which density fluctuations are suppressed. The
phase-fluctuations then decay to produce the true phase-coherent
condensate, with a characteristic time scale $\tau_\phi$ which
increases with the system size $L$: $\tau_\phi\propto L$ in the
collisionless  regime and $\tau_\phi\propto L^2$ in the
hydrodynamic regime. Although our trapped system differs from the
homogeneous system considered in Ref.~\cite{Kag92}, Svistunov
\cite{Svi01} points out that this theory can be applied to trapped
hydrodynamic clouds, where the trapping potential can be
neglected.  In this case, the resulting quasi-condensate will be
out-of-equilibrium with respect to a global coherent motion in the
trap, thereby exciting a quadrupole mode similar to that observed
in Refs.~\cite{Shv02, Ric03}. Indeed, for a higher atom number
$N_{\rm{i}}=8.6\times 10^5$ we directly observe quadrupole
oscillations, with an amplitude (in the trap)  of $12\,\mu$m and a
decay constant of about 250\,ms. Oscillations of the axial
condensate length broaden the momentum distribution, thus the
excess momentum widths in Fig.~\ref{fig:Delta_nu_comp} can be
attributed to quadrupole oscillations with amplitudes of $4\,\mu$m
($N_{\rm{i}}=3.8\times 10^5$) and $5.5\,\mu$m
($N_{\rm{i}}=7.2\times 10^5$) \cite{amp} and decay constants of
about 700\,ms and 300\,ms respectively.   These values are
consistent with those obtained for $N_{\rm{i}}=8.6\times 10^5$,
assuming an oscillation amplitude and decay rate which increase
with the atom number.  Therefore, apart from this decaying
quadrupole mode, we conclude that higher-order, non-equilibrium
phase fluctuations have decayed within a time shorter than 100\,ms
after the onset of condensation, in qualitative agreement with the
predictions of Kagan \textit{et al.} \cite{predictions}.

This result allows us to interpret the data of
Fig.~\ref{GrowthCurves}, where the observed growth is slower than
that predicted for a phase-coherent condensate.  We suggest a
simple picture of a 1D phase-fluctuating condensate as being made
up of a number of phase-coherent domains. As the growth rate
includes a factor $(1+N_0)$ due to bosonic stimulation, the growth
rate of each domain will be smaller than that of a single large
condensate, thereby reducing the overall growth rate. We have
shown that the coherence length of our condensate approximately
follows the equilibrium prediction, $L_\phi$, which can be
calculated using the instantaneous values of $N_0$, $L$ and $T$
measured during the observed growth.  We therefore modify the
growth rate in the simulation, replacing the factor $1+N_0$ with
$1+L_\phi N_0/L$ \cite{ignore_osc}. This slows the simulated
growth rate, obtaining improved agreement with the data as shown
by the solid lines in Fig.~\ref{GrowthCurves} \cite{rescale}.
However, the added delay time remains unexplained.

In conclusion, we have observed the growth of a condensate in an
elongated trap and have shown, by comparison with numerical
simulations, that the presence of phase fluctuations slows the
growth of the condensed fraction. By studying the evolution of the
momentum width during condensate growth, we have directly observed
the growth of the phase coherence with time. Compared with that
expected for a condensate at equilibrium, these measurements
reveal a broadening of the momentum distribution during growth,
compatible with quadrupole shape oscillations. Apart from this
decaying oscillation, we conclude that the condensate has already
reached the equivalent equilibrium coherence length within 100\,ms
after the onset of condensation. Because a complete theory for our
experimental situation is lacking, the agreement between theory
and experiment can only be qualitative and an extension of the
model of Ref.~\cite{Kag92} to trapped condensates, particularly in
this quasi-1D geometry \cite{Prou05}, is required.

Using the equilibrium coherence length, we accurately predict the
slowing of the growth rate due to phase fluctuations. However,
there remains a discrepancy on the delay time before condensation,
which may indicate non-equilibrium excitations at early times, as
predicted by Ref.~\cite{Kag92}. A measurement of the phase
coherence length at shorter times is needed.  This is exceedingly
difficult in our system, since at short times the condensed
fraction is too small to obtain clear Bragg spectra. Other
techniques might be used instead, such as atom laser correlation
measurements \cite{Ess1}, combined with single-atom detection
\cite{Ess2,Wes05}.

We thank L.~Sanchez-Palencia,  G.~V.~Shlyapnikov, F.~Gerbier, S.
Richard and J.~H.~Thywissen for useful discussions. We acknowledge
support from IXSEA (M.H.), the Marie Curie Fellowship Programme
(J.R.), the Fundaci\'{o}n Mazda para el Arte y la Ciencia (A.V),
the Australian Research Council (M.D.), the  D\'{e}l\'{e}\-gation
G\'{e}n\'{e}rale de l'Armement, the Minist\`ere de la Recherche
(ACI Nanoscience 201), the European Union (grants IST-2001-38863
and MRTN-CT-2003-505032) and the ESF (QUDEDIS programme).  The
atom optics group of the Laboratoire Charles Fabry is a member of
IFRAF (www.ifraf.org).

%****************************************************************


\begin{thebibliography}{Shv02}
\bibitem{hydro}{For the thermal cloud immediately before the
shock-cooling ramp, we obtain $0.04<\omega_z
\tau_{\rm{coll}}<0.13$.}
\bibitem{rescale}{All raw measured atom numbers $N$ have been rescaled by
a factor 0.9 (consistent with our 20\% estimated  error) in order
to obtain better agreement between experiment and theory.}
\bibitem{amp}{For such small amplitudes, the oscillations cannot be directly
observed in the time-of-flight images due to the limited
resolution of our imaging system.}
\bibitem{predictions}{With our experimental parameters we calculate from \cite{Kag92} a damping time $\tau_\phi$ between 10\,ms and 500\,ms, depending on the collisional regime.}
\bibitem{ignore_osc}{Here, we ignore the additional quadrupole
oscillations.}
\end{thebibliography}
\end{document}